\documentstyle[12pt,epsfig]{article}
\newcommand {\be}{\begin{equation}}
\newcommand {\ee}{\end{equation}}
\newcommand {\bey}{\begin{eqnarray}}
\newcommand {\eey}{\end{eqnarray}}

\begin{document}
\pagestyle{empty}

\title{Lyapunov exponents from node-counting arguments}

\author{
Antonio Politi$^{1}$\thanks{also Istituto Nazionale di Fisica Nucleare, 
Sezione di Firenze, I-50125 Firenze, Italy }
Alessandro Torcini$^{2}$\thanks{also Istituto Nazionale di Fisica della Materia, 
Unit\`a di Firenze, I-50125 Firenze, Italy }
and Stefano Lepri$^{3}$ \\
$^1$ {\small \it Istituto Nazionale di Ottica I-50125 Firenze, Italy}\\
$^2$ {\small \it Dipartimento di Energetica ``S. Stecco'' I-50139 Firenze, 
Italy}\\
$^3$ {\small \it Max-Planck-Institut f\"ur Physik komplexer Systeme
D-01187 Dresden, Germany} \\
}
\maketitle
\begin{abstract}
A conjecture connecting Lyapunov exponents of coupled map lattices
and the node theorem is presented. It is based on the analogy between
the linear stability analysis of extended chaotic states and the 
Schr\"odinger problem for a particle in a disordered potential. As a
consequence, we propose an alternative method to compute the Lyapunov 
spectrum.  The implications on the foundation of the recently proposed 
``chronotopic approach'' are also discussed.
\end{abstract}

\section{INTRODUCTION}

In a series of recent papers \cite{lyap1,lyap2,lyap3}, the so-called 
{\it chronotopic 
approach} has been developed with the aim of extending the by-now standard 
concept of Lyapunov spectrum \cite{ruelle} to spatially inhomogeneous 
perturbations. In 1d extendend systems, the study of infinitesimal 
perturbations with an exponential profile (and a generic decay-rate $\mu$) 
has led to introduce the generalized integrated density
$n_\lambda(\lambda,\mu)$, defined as the fraction of Lyapunov exponents
smaller than $\lambda$. In these notations, the usual spectrum is denoted by
$n_\lambda(\lambda,0)$, as it is characterized by the absence of an overall 
exponential spatial profile. In a complementary way, the spatial dynamics of 
perturbations has been studied by assigning a temporal growth rate $\lambda$
and thereby determining $\mu$, by following the evolution in tangent space 
along the space axis.  In this case, one determines the spectrum of spatial
exponents $n_\mu(\lambda,\mu)$. The main result
of the chronotopic approach is the existence of a dynamical invariant, the 
{\it entropy potential} $\Phi$, the knowledge of which allows to determine all
properties of the evolution of localized as well as extended perturbations.
However, the existence of $\Phi$ has been proved only in a few, very simple,
cases of uniform space-time patterns and numerically tested just in some 
more realistic examples. No general argument has yet been found to justify 
the validity of the whole approach in generic 1d systems. The available
``proof'' (see in particular Ref.~\cite{lyap2}) is essentially
based on the observation that Lyapunov vectors (i.e. the eigenvectors of the
stability matrix) are ordered from the most to the least unstable one (or
viceversa) for increasing wavenumber $k$. Accordingly, one can describe 
the spatial structure of a generic Lyapunov vector with a single complex
number $\tilde \mu = \mu + i k$, the real part of which is the exponential
growth
rate, while the imaginary part is nothing but the wavenumber or,
equivalently, the integrated density $n_\lambda(\lambda,\mu)$. Analogously, a 
temporal frequency $\omega$ has been invoked to order all spatial Lyapunov 
exponents and thus to represent a measure of $n_\mu(\lambda,\mu)$. The 
frequency $\omega$ can be read as the imaginary part of the complex number
$\tilde \lambda = \omega + i \lambda$, where $\lambda$ is the
temporal growth rate (i.e. the Lyapunov exponent) of the given perturbation.
The analyticity properties of the ``dispersion relation'' connecting 
$\tilde \mu$ with $\tilde \lambda$ furnish the last ingredient to ``prove'' 
the existence of an entropy potential \cite{lyap2}. The key question is 
how far can one go with this type of arguments to prove the validity of the 
chronotopic approach?
If we look at one particular consequence of the existence of $\Phi$, namely
that the Kolmogorov-Sinai entropy density $h_{KS}$ is independent of the 
direction along which a 2d space-time pattern is thought to be generated, 
then we are led to expect a rather general validity. In fact, it looks 
rather plausible that $h_{KS}$ is an intrinsic property of a given pattern,
independent of the way we look at it!

In this paper we explore the possibility to introduce a general but 
meaningful definition of the wavenumber and, in turn, to define 
``rotation numbers'' as the imaginary counterpart of the Lyapunov exponents.
As a result, we propose an alternative method to compute the
Lyapunov spectrum by using the transfer matrix approach rather than
iterating the linear relations in time as it is usually done. The approach
is limited to a class of coupled map lattices (CMLs) with everywhere expanding 
multipliers. Accordingly, it is not yet demonstrated that
the perspectives so far outlined are consistent in general. Nevertheless, our
results provide encouraging indications for future investigations.

The outline of the paper is as follows. In section 2, we discuss the
simple case of fixed points (in time). In section 3, the approach is extended
to periodic and chaotic patterns, introducing a conjecture and numerically
testing it. In the last section, we discuss the limitations as well as 
possible further extensions.

\section{\bf ANALOGY WITH THE SCHR{\"O}DINGER PROBLEM}

The computation of both temporal and spatial Lyapunov spectra is normally
carried out by resorting to the standard orthonormalization procedure (SOP)
introduced many years ago \cite{benettin}. In this section, we discuss the
analogy between the linear stability analysis of CMLs and the 1d
Schr\"odinger problem (see also \cite{paladin}), with the aim of both
strenghtening the internal consistency of the chronotopic approach and to
introduce the first elements of an alternative algorithm.

Let us consider a CML \cite{cml} and denote with $x_n^i$ the field variable
at lattice location $i$ ($i=1,\dots,L$) and time $n$. By introducing the
$L$-dimensional column vector $\delta X_n$ of the perturbations $\delta
x_n^i$, we can synthetically express the evolution equations in the tangent
space as
\be
\delta X_{n+1} \;=\; M_n \, D_\varepsilon \, \delta X_n \quad ,
\label{tgspace}
\ee
where $M_n$ is a diagonal matrix whose diagonal elements $m^i_n$ are the 
derivatives of the (here unspecified) nonlinear mapping, and $D_\varepsilon$
is the tridiagonal matrix associated to the diffusive coupling 
\be
\left(D_\varepsilon\right)_{i,j} \;=\; 
{\varepsilon\over 2} \, \delta_{i+1,j} +
(1- \varepsilon) \, \delta _{i,j} +
{\varepsilon\over 2} \, \delta_{i-1,j} \quad .
\ee
A simple but instructive example that allows discussing the main ideas is
that of frozen random patterns, for which $m^i_n$ depends only on the
spatial variable $i$. In this case, the estimation of the Lyapunov exponents
$\lambda$ reduces to the eigenvalue problem for the matrix $MD_\varepsilon$,
namely 
\be 
\Lambda \, \delta x^i = m^i \bigg[ {\varepsilon\over 2} \, \delta x^{i-1}
    + (1- \varepsilon)\, \delta x^i + {\varepsilon \over 2}\, \delta x^{i+1}
\bigg]
\label{eig}
\ee
where $\lambda = \log |\Lambda|$.

Eq.~(\ref{eig}) resembles the tight-binding approximation of the 1d
Schr\"odinger equation (with imaginary time) in the presence of a random
potential $V^i$, namely the celebrated Anderson model
\be
\omega \, \psi^i = \psi^{i+1} + \psi^{i-1} + (V^i-2) \psi^i \quad ,
\label{ander}
\ee
where the eigenvalue $\omega$ plays the role of the multiplier $\Lambda$,
while the eigenfuction corresponds to the Lyapunov vector of the CML.
Accordingly, finding the spectrum of the Schr\"odinger operator is
equivalent to finding the Lyapunov spectrum of the CML \cite{pla}.
Incidentally, notice that an even closer analogy exists with the
computation of the vibrational spectrum of a chain with random masses
\cite{mattis}.

The spectrum (or the density of states) of the Schr\"odinger problem can be
determined without actually diagonalizing the operator implicitely defined
by the r.h.s. of Eq.~(\ref{ander}) (which is just the sum of the discretized
Laplacian and a diagonal operator). Indeed, its symmetry ensures the
validity of the node theorem which states that the eigenfunctions are
ordered according to the number of their zeros \cite{pastur}. Accordingly,
the structure of a given eigenfunction suffices to determine the position of
the corresponding eigenvalue inside the energy spectrum. This task is
usually accomplished by the transfer matrix approach.

It is natural to ask whether $n_\lambda$ in a CML can be analogously
determined from the spatial structure of the corresponding Lyapunov vector.
This question can be easily answered in the positive for frozen random
patterns. In fact, although the operator in the r.h.s. of Eq.~(\ref{eig}) is
not symmetric, one can easily realize that the change of variables 
$\delta x^i \to \sqrt{m^i} \,\delta x^i$, leads to a fully symmetric
structure. As a consequence, the node theorem applies also in this restricted
CML problem. Furthermore, the symmetry of the operator guarantees that the
eigenvalues are all real, i.e. no rotations in tangent space are involved.

In practice, it is sufficient to proceed as follows: one starts fixing
$\Lambda = \exp(\lambda)$ in Eq.~(\ref{eig}), where $\lambda$ is the Lyapunov
exponent of interest. We then iterate Eq.~(\ref{eig}) along a spatial
direction (left and right are equivalent directions) and count the number of
changes of sign of $\delta x^i$ from one to the next site. The fraction of
such changes (i.e. the number of zeros per unit length) equals the
integrated density of Lyapunov exponents $n_\lambda$. In Fig.~1 we compare
the results of such a procedure with the SOP. The perfect agreement
represents a direct verification of the validity of the approach.

\begin{figure}[htbp]
\centering\epsfig{file=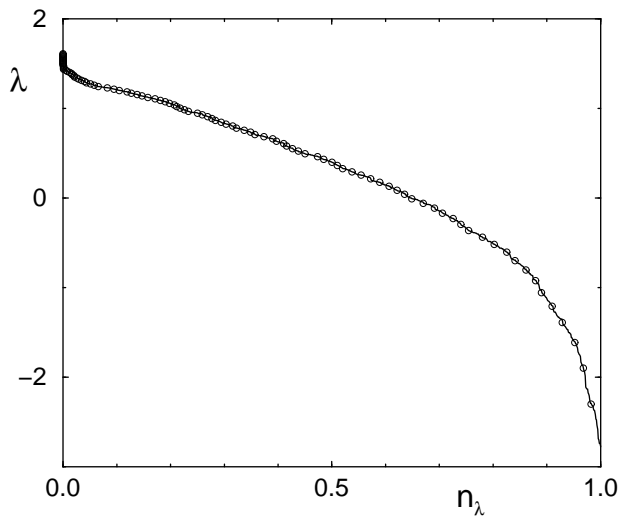,angle=-90,width=0.45\linewidth}
\caption{Lyapunov spectrum for a frozen random pattern. The solid curve
refers to the result of the SOP, while open circles correspond to the
outcome of the node counting approach. Here and in all subsequent cases,
we have always considered $\varepsilon = 1/3$.}
\end{figure}

Let us notice that, from the point of view of the chronotopic approach,
this way of determining $n_\lambda$ is very close to the method used to
define the spatial spectrum $n_\mu$, the main difference being that, instead 
of computing the growth rates, we look at the nodes, i.e. at the spatial 
frequency or ``average wavenumber''  $k$ of the Lyapunov vector.

\section{THE GENERAL CASE}

In this section we extend to time-dependent patterns the ideas sketched
above for stationary random trajectories. More specifically, we
shall consider orbits of temporal period $T$ with $T>1$. In this case, we
have to solve the eigenvalue problem for the product matrix
\be
U = \prod_{j=1}^T M_j D_\varepsilon \quad .
\label{matrix}
\ee
Notice that $U$ is a banded matrix (of band width $2T+1$) so that we
are dealing with a sort of Schr\"odinger problem with long-range hopping.
The fundamental difference is that not only $U$ is not symmetric, but
no similarity transformation can turn it into a symmetric matrix. This is
confirmed by the generic existence of complex eigenvalues for $T>2$
\footnote{For period-2 orbits it is still possible to reduce the matrix to
a symmetric form with a suitable change of variables.}.

It has to be admitted that this represents a serious mathematical
difficulty, as the node theorem is rigorously proved (at least to our
knowledge) only for a class of operators with a strictly real and positive
spectrum \cite{gant}. Nevertheless, one can at least hope that some sort of
ordering is maintained in the case of Lyapunov exponents.
In fact, the latter are the logarithms of the eigenvalues of a  
matrix which is the the product of $U$ by its transpose. 
Such a matrix is clearly symmetric and has a real and positive spectrum
(we assume that no zero eigenvalues are present).

Nonetheless, even having accepted this optimistic point of view, one has to
face further difficulties. The straightforward generalization of the method
of the previous section would amount first to rewriting (\ref{tgspace}) as a
spatial mapping. This requires the knowledge of the variables $\delta x^i_n$
at all times in two consecutive sites and, accordingly, reads as
\be
\delta Y^{i+1} = L^i(\Lambda) \, \delta Y^i
\label{form}
\ee
where $\delta Y$ is now a vector consisting of $2T$ components and
$L^i(\Lambda)$ is the $2T\times 2T$ transfer matrix.

At variance with the case $T=1$ discussed in the previous section, we cannot
expect to get the correct value of $n_\lambda$ by simply counting the nodes
of one of the $\delta x^i_n$ resulting from the repeated application of
$L^i$ to a randomly chosen initial vector $\delta Y^0$. In fact, the single
eigenvector associated to a given eigenvalue (if we disregard the unlikely
occurrence of degeneracies) corresponds to a unique trajectory of the
$2T$-dimensional transformation $L^i$ (apart from an irrelevant scaling
factor). Accordingly, it is very unlikely that a random choice of the initial
conditions yields the right spatial structure, unless the number of nodes is
independent of the trajectory, a possibility that must be ruled out on the
basis of our numerical studies. This problem is very much reminiscent of the
difficulty to determine the standard Lyapunov spectrum: in order to compute
the $m$-th Lyapunov exponent, one cannot start from a randomly chosen
initial condition: it is necessary to select $m$ linearly independent
vectors \cite{benettin}.

Notice that this very same problem would occur also in the safer case
where node counting is known to apply, such as the Anderson model with
long-range hopping or the harmonic chain with next-to-nearest neighbours
coupling. Actually, an extension of the method based on counting the
sign-changes of principal minors has been devised in the literature
\cite{martin}, but it seems definitely unpractical for matrices of large
bandwith (say for $T>3$).

Having recognized such difficulties, and inspired by the analogy with the
SOP, we have looked for a similar procedure in the present context,
eventually finding an approach that works in all cases we have considered.
As we have been unable to develop a rigorous proof, we present it here as a
\vskip .5cm
{\it Conjecture: The integrated density of Lyapunov exponents 
$n_\lambda(\lambda,0)$ for a periodic orbit of period $T$ coincides with 
the density of nodes along the $T$-th most expanding direction of the product 
of transfer matrices $L^i({\rm  e}^\lambda)$.}
\vskip .5cm
In order to make the above conjecture really transparent, we need first to
define the $m$-th expanding direction $Z^i_{(m)}$. We know that a generic
vector $V^i_{(1)}$ aligns, after a suitable transient, along the most
expanding direction $Z^i_{(1)}$, which can thus be easily identified. The
concept of $m$-th most expanding direction is not, however, equally clear.
In fact, we know that if we take $m$ independent vectors $V^i_{(1)},\ldots
V^i_{(m)}$, the subspace $S^i(m)$ identified by the $m$ vectors
asymptotically orients itself as the most expanding $m$-dimensional subspace
\cite{benettin}, but there is not a unique basis which identifies a given
subspace $S^i(m)$: any set of linearly independent vectors generating
$S^i(m)$ is equally meaningful. In fact, the SOP exploits this freedom to
determine the vectors by imposing the mutual orthogonality, a condition
motivated by the opportunity to minimize the numerical error. However, we
have verified in several cases that this somehow arbitrary choice is not
appropriate for the present purpose.

\begin{figure}[htbp]
\centering\epsfig{file=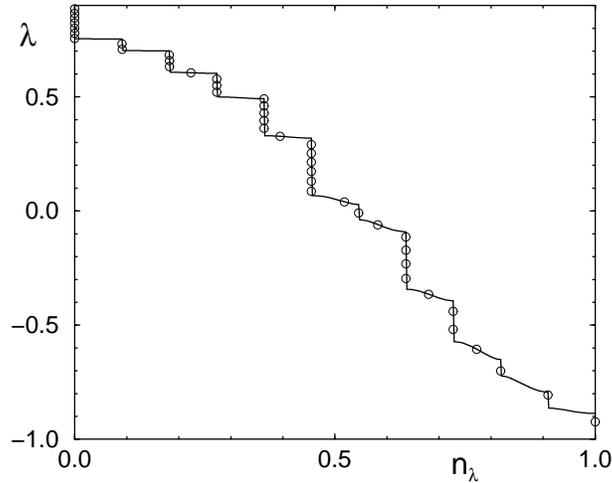,angle=-90,width=0.45\linewidth}
\caption
{Lyapunov spectrum for an orbit of spatial period 11 and temporal period
$T=7$. The solid curve refers to the SOP, while open circles are the outcome
of the node counting approach.}
\end{figure}

A meaningful solution can be found by realizing that $Z^i_{(m)}$ is the
least expanding direction in the subspace $S^i(m)$. As a consequence,
$Z^i_{(m)}$ is also the most expanding according to the backward evolution
in $S^i(m)$. Since the most expanding direction of a given mapping is the
only one which can be directly identified, we have finally an algorithm to
determine $Z^i_{(m)}$, an algorithm that can be also taken as an operative
definition of $m$-th most expanding direction.
More precisely, one first iterates $m$ vectors according to the general
relation (\ref{form}) in order to determine the sequence of subspaces
$S^i(m)$ and the rules for the mapping of $S^i(m)$ onto $S^{i+1}(m)$.
Afterwards, one must iterate a generic vector backward in space,
restricting the dynamics to the sequence of subspaces $S^i(m)$. Notice that
this restriction is very important, since any direction $Z^i_{(j)}$ with
$j > m$ is more unstable than $Z^i_{(m)}$ so that any small but unavoidable
numerical error would soon drive the trajectory towards the most unstable
direction in the whole space $R^{2T}$.

Once we have been able to identify the $m$-th expanding direction, we can
compute the nodes of $\delta x^i_n$ along the $T$th direction and thus
test the conjecture about $n_\lambda(\lambda,0)$. In Figs.~2 and 3 we report
two of the many examples we have studied to compare the Lyapunov spectrum
determined through the SOP (solid curves) with the outcome of the node
counting approach (open circles). In all cases we have found that the
agreement is within the numerical accuracy.

\begin{figure}[h]
\centering\epsfig{file=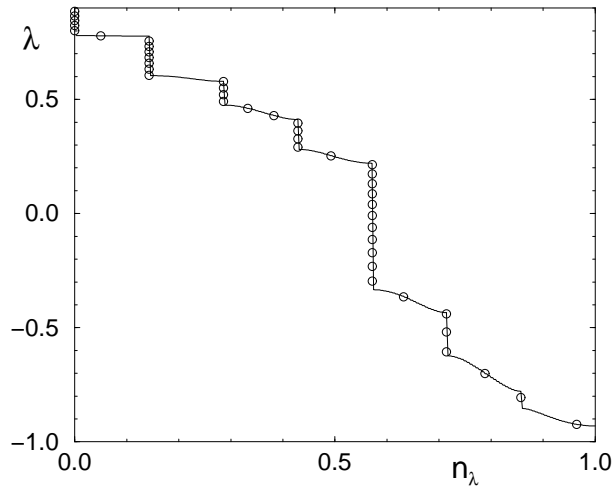,angle=-90,width=0.45\linewidth}
\caption{Same as in Fig.~2 for an orbit of spatial period 7 and temporal
period $T=5$.}
\end{figure}

\begin{figure}[htbp]
\centering\epsfig{file=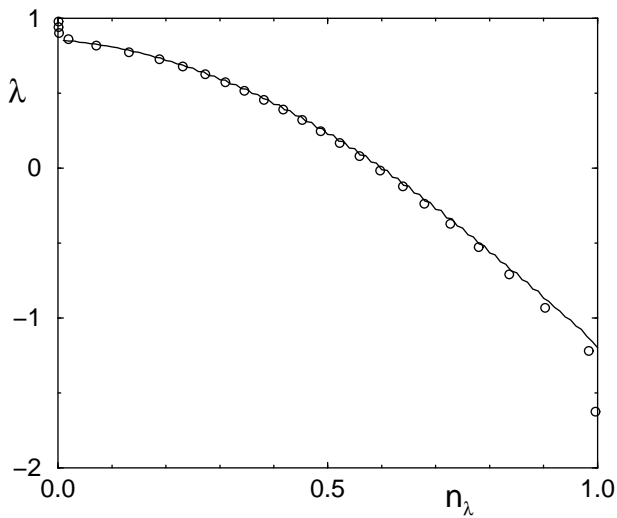,angle=-90,width=0.45\linewidth}
\caption{Lyapunov spectrum for a spatio-temporal random pattern. As for
the above cases, open circles follow from the node counting while the
solid curve correspond to the SOP.}
\end{figure}

It is important to stress that the conjecture appears to hold also in the
fully chaotic regime, i.e. for random patterns both along the spatial and
temporal direction. In this case, one has formally to consider temporal
stripes of increasing height, i.e. to take the limit $T \to \infty$. The
results reported in Fig.~4 for one such case hint at a general validity of
the correspondence between node counting and Lyapunov spectra.

\section{CONCLUSIONS AND PERSPECTIVES}

In the previous section, we have seen that the spatial structure of
the Lyapunov vector corresponding to the exponent $\lambda$ contains the
information necessary to determine the integrated density without the need
to consider all Lyapunov exponents larger than $\lambda$, as required by
the SOP. Although this may be considered as a computational advantage,  
we want to emphasize that the relevance of our conclusions,
where proved to be rigorously true, does not come from the opportunity
offered by the new algorithm.
Indeed, we have seen that the study of the evolution along the spatial
direction is not logically different from the application of the SOP used to
follow the temporal evolution. The difference is that the size of the
matrices involved in the evolution in the tangent space does not depend
on the size of the system but on the periodicity of the solution. In the
generic case of space-time chaos, it is a matter of the convergence rate
versus space and time that makes one approach preferable to the other
\cite{arka}.

The relevance of our result relies on the possibility to attribute a meaning
to the average spatial frequency of the Lyapunov vectors (the fraction of
nodes). This confirms the intuition that not only the real but also the
imaginary parts of the expansion rates $\tilde \mu$ are meaningful
quantities and both contribute to the validity of the chronotopic approach.
Let us indeed recall that the only cases in which we have been able to prove
the existence of an entropy potential are those in which we have been able to
interpret the spatial (and temporal) frequencies as suitable integrated
densities.

The apparently coherent link between the node counting and the chronotopic
approach is testified by the following extension of the conjecture in
Sec.~3:

{\it The integrated density of Lyapunov exponents $n_\lambda(\lambda,\mu)$
coincides, for $T$ large enough,  with the density of nodes along the $m$-th
most expanding direction of the product of transfer matrices 
$L^i({\rm e}^\lambda)$, where $m$ is fixed by the implicit condition 
$n_\mu(\lambda,\mu)=m/T$}.

The above conjecture states that the nodes have to be counted along the
direction characterized by the preassigned spatial growth rate $\mu$.
In practice, we fix $\lambda$ (a free parameter in the transfer matrix
expression) and iterate $m$ vectors to determine the $m$-th most
expanding direction.
The rate $\mu$ is the spatial Lyapunov exponent corresponding to the
$m$-th direction, while $n_\lambda$ is estimated by counting the corresponding
nodes.
The validity of the result can be tested by imposing a spatial profile
equal to ${\rm e}^\mu$ in the temporal evolution and 
thereby determine $n_\lambda$ and $\lambda$ with the SOP 
(this last value is the one reported in parenthesis in the table). 
In all cases we have considered, we have always found a
good agreement. The data reported in the table refer to a uniform
distribution of multipliers between 0 and 5.

A serious limitation to the validity of the results discussed
in this paper is the restriction to positive multipliers (i.e. a positive
slope of the local map, as for the generalized Bernoulli shift): all
results have indeed been obtained for strictly positive $m^i_n$. However,
we are confident that this limitation can be lifted and we are indeed
working to clarify this crucial point.

\begin{table}[h]
\caption[tabone]
{Comparison of the temporal Lyapunov spectrum $\lambda = \lambda(\mu)$ 
as determined with the
node counting approach and the SOP (the value in parenthesis),
for various combinations  of the quantities $n_\mu$, $\mu$ and $n_\lambda$.
In the last column, the relative error (\%) is reported.}
\vskip .5 truecm
\centerline{
\begin{tabular}{rrrrr}
\hline
\hline
\hfil $n_\mu$ \hfil \hfil & \hfil $\mu$ \hfil 
& \hfil $n_\lambda$ \hfil &
${\rm e}^\lambda$ \hfil \hfil & \hfil \hfil \hfil error  \\
\hline
\hline
0.0 &  1.711  & 0.005 & 4.000 (3.989) & 0.2 \\
0.5 &  2.635  & 0.602 & 4.000 (4.032) & 0.8 \\
0.9 &  3.122  & 0.924 & 4.000 (4.028) & 0.7 \\
\hline
\end{tabular}
}
\end{table}

More important, in our opinion, is the question whether the same approach
can be extended to continuous-time and -space systems, i.e. to more realistic
models of space-time chaos. We believe that instead of checking numerically
whether this is true or not, it is more important to look for the possibly
deep reasons that lie behind the apparent validity of the conjectures
presented in this paper.

\section*{Acknowledgments}

We acknowledge the hospitality of ISI-Torino during the activity of the EC
Network CHRX-CT94-0546.

\end{document}